\documentclass[%
preprint,
 amsmath,amssymb,
 aps,
]{revtex4-1}

\usepackage{graphicx}
\usepackage{dcolumn}
\usepackage{bm}

\begin{document}


\title{
Unitary Representation of Symplectic Group \\
for Phase Point Operators on Discrete Phase Space
}

\author{D. Watanabe}
 \email{watanabe@rbi.apphy.u-fukui.ac.jp}
\author{T. Hashimoto}%
 \email{hasimoto@u-fukui.ac.jp}
\author{M. Horibe}%
 \email{horibe@u-fukui.ac.jp}
\author{A. Hayashi}%
 \email{hayashia@u-fukui.ac.jp}
\affiliation{%
Department of Applied Physics, University of Fukui \\
Bunkyo 3-9-1, Fukui 910-8507, Japan
}


\date{\today}

\begin{abstract}
The phase point operator \(\Delta(q,p)\) is the quantum mechanical counterpart
of the classical phase point \((q,p)\).
The discrete form of \(\Delta(q,p)\) was formulated
for an odd number of lattice points by Cohendet et al.
and for an even number of lattice points by Leonhardt.
Both versions have symplectic covariance,
which is of fundamental importance in quantum mechanics.
However, an explicit form of the projective representation of the symplectic group
that appears in the covariance relation is not yet known.
We show in this paper the existence and uniqueness of the representation,
and describe a method to construct it using the Euclidean algorithm.
\end{abstract}

\maketitle

\section{Introduction}

The Wigner function was introduced by Wigner
and utilized to study the quantum correction for thermodynamics in 1932 \cite{Wigner}. 
In recent years, its range of applications has extended to
quantum optics, quantum chaos, quantum computing, and other fields,
and it has again become a focus of interest for research
in which quantum-classical correspondence is essential.
The history of the Wigner function on discrete phase space is
relatively young and marked in particular by its application
to discrete phase space composed of a prime number of lattice points (prime-lattice phase space),
formalized by Wootters in 1987 \cite{Wooters},
and to discrete phase space composed of an odd number of lattice points (odd-lattice phase space)
corresponding to integer spin, formalized by Cohendet et al. in the same year \cite{Cohendet}. 
However, it was pointed out that its behavior on discrete phase space
composed of an even number lattice points (even-lattice phase space)
was found to differ
substantially from that on odd-lattice phase space.
In 1995, Leonhardt formulated the Wigner function on even-lattice phase space
corresponding to half-integral spin, but found it necessary to incorporate
a virtual degree of freedom (so-called ghost variable) \cite{Leonhardt1,Leonhardt2}.
The Wigner function becomes a distribution of position
upon integration over all momentum space.
Conversely, integration over position transforms the Wigner function
into a momentum distribution, 
and thus it exhibits the behavior of a distribution function. 
However, the values can be negative and it is therefore called a quasi-distribution function.
Various functions have this marginality in general, but it is known that
the Wigner function provides the unique solution having rotational symmetry on continuous space. 
Symplectic transformation
yields an invariant canonical commutation relation
and is therefore an important symmetry in quantum mechanics.
It is known that on continuous or odd-lattice phase space,
if the phase point operator is sandwiched between a Fourier operator and its Hermitian conjugate,
the argument of the operator rotates 90 degrees.
%
%
It represents the simplest symplectic covariance (Fourier covariance)
among linear canonical transformations.
In the present article, we show that on odd- and even-lattice phase spaces
the phase point operator derived from the Wigner function by Cohendet et al. and Leonhardt
has symplectic covariance and that a projective representation of
such a symplectic transformation group exists and is unique. 
We discuss the fundamentals of phase point operators on discrete phase space in Sec. II to IV,
describe symplectic covariance in Sec. V, and define the symplectic transformation group
necessary for symplectic covariance on discrete phase space and show that they are formed by two elements
(\(h_+\) and \(h_-\)) in Sec. VI.
In Sec. VII, we prove that the unitary projective representation \(U(S)\) of
symplectic group exists and that it is unique.
In Sec. VIII, we seek \(U(h_+)\) and \(U(h_-)\) on low dimensional odd-lattice phase space
from the two elements shown in Sec. VII based on the necessary conditions,
predict the forms for general odd-number dimensions,
and then confirm that the predicted forms have symplectic covariance
(confirmation of sufficiency).
In Sec. IX, we similarly seek \(U(h_+)\) and \(U(h_-)\) on even-lattice phase spaces
based on the necessary conditions, but find that they do not have a projective representation
with the symplectic transformation group defined in Sec.VI. 
We therefore consider anew the use of the virtual degree of freedom (ghost variable),
redefine the symplectic transformation group with the number of lattice points doubled,
and find \(U(h_+)\) and \(U(h_-)\) that have symplectic covariance for general even-number dimensions.

\section{Wigner function on continuous phase space}

\subsection{Definition and marginal property}

The Wigner function, which was originally a function on classical continuous phase space,
is defined as
\begin{eqnarray}
{\cal W}(q,p)={1\over2\pi\hbar}\int_{-\infty}^\infty dr e^{ipr/\hbar}\psi^*(q+{r\over2})\psi(q-{r\over2})
.
\end{eqnarray}
Integration over momentum and position gives
\begin{eqnarray}
\int_{-\infty}^\infty{\cal W}(q,p)dp
&=&|\psi(q)|^2
,\\
\int_{-\infty}^\infty{\cal W}(q,p)dq
&=&|\tilde{\psi}(p)|^2
,
\end{eqnarray}
respectively. 
It thus becomes a position distribution when integrated over momentum and a momentum distribution
when integrated over position.
The two wave functions \(\psi(q)\) and \(\tilde\psi(p)\) are interconvertible via Fourier transformation.
As evident from this marginal property (we refer to it as marginality), the Wigner function is a kind of quantum-mechanical distribution function,
but it yields negative values and is therefore referred to as a quasi-distribution function. 

\subsection{The phase point operator}

The phase point operator \(\Delta(q, p)\) is defined as the state independent part
of the Wigner function,
\begin{eqnarray}
{\cal W}(q,p)={1\over2\pi\hbar}{\rm Tr}[\rho\Delta(q,p)]\ \ \ ,\ \ \ \rho=|\psi\rangle\langle\psi|
.
\end{eqnarray}
In the position representation it is given by 
\begin{eqnarray}
\Delta(q,p)=\int_{-\infty}^\infty dr e^{ipr/\hbar}|q+{r\over2}\rangle\langle q-{r\over2}|\label{fano1}
,
\end{eqnarray}
and in the momentum representation by
\begin{eqnarray}
\Delta(q,p)=\int_{-\infty}^\infty ds e^{-iqs/\hbar} |p+{s\over2} \rangle \langle p-{s\over2}|
.
\end{eqnarray}
With the phase point operator, a classical Hamiltonian can be transformed
to a quantized one \(\hat{\cal H}\) with Weyl ordering:
\begin{eqnarray}
{\hat{\cal H}}_{\rm Weyl}({\hat q}, {\hat p})
=
\frac1{2\pi\hbar}
\int_{-\infty}^\infty \int_{-\infty}^\infty dq\:dp\ {\cal H}(q,p) \Delta (q,p)  
.
\end{eqnarray}
The phase point operator \(\Delta(q, p)\) can therefore be regarded as a quantum operator
corresponding to the classical phase point \((q, p)\). 
From the marginality of the Wigner function, it has the following properties,
\begin{eqnarray}
\frac1{2\pi\hbar}
\int_{-\infty}^\infty dp \Delta (q,p)&=&| q \rangle\langle q|, \\
\frac1{2\pi\hbar}
\int_{-\infty}^\infty dq \Delta (q,p)&=&| p \rangle\langle p|,
\end{eqnarray}
which are called the operator form of the marginality.
An important advantage of considering the phase point operator is its capability for quantization of
various geometrical objects
\cite{
Przanowski1,
Przanowski2},
especially discrete systems
\cite{
Santhanam1,
Santhanam2,
Santhanam3,
Jagannathan,
Ruzzi1,
Ruzzi2,
Chaturvedi1,
Galetti,
Marchiolli1,
Marchiolli2,
Marchiolli3,
Chaturvedi2}.
We refer to Ref.
\cite{
Stovicek,
Vourdas,
Chaturvedi3,
Mukunda}
for quantum mechanics on finite abelian or Lie groups.
\\
%

\section{Wigner function by Cohendet et al. and the phase point/Weyl operators on odd-lattice phase space}

Beginning this section with the Weyl operator formulated by Cohendet et al.,
we briefly overview the phase point operator on discrete phase space
composed of an odd number of lattice points (odd-lattice phase space).

\subsection{Definition of the Weyl operator in odd-lattice case}

Cohendet et al., in composing the Wigner function on odd-lattice phase space,
first defined the Weyl operator as
\begin{eqnarray}
({W^{C}}_{m,n}\psi)(k) = \exp\left({-\frac{4\pi imn}{N}}+{\frac{4\pi ink}{N}}\right)
\times \psi(k-2m),
\end{eqnarray}
where \(m,n,k\in I=\{-\frac{N-1}2,-\frac{N-3}2,\cdots,\frac{N-3}2,\frac{N-1}2\}\) and \(N\) is an odd integer.
The integer \(N\) is regarded as the modulus in \(I\).


%
The phase \(Q\), shift \(P\), and inversion \(T\) operators are defined as follows
for convenience in calculation,
\begin{eqnarray}
Q=\sum_{k}|k \rangle \omega^{k}\langle k| \label{Cohendet1},
\end{eqnarray}
\begin{eqnarray}
P=\sum_{k}|k-1\rangle \langle k| \label{Cohendet2}
,
\end{eqnarray}
\begin{eqnarray}
T=\sum_{k}|-k\rangle \langle k| \label{Cohendet3}
,
\end{eqnarray}
where \(\omega\) is the primitive \(N\)-th root of unity:
\begin{eqnarray}
\omega = \exp\left({2\pi i\over N}\right)
.
\end{eqnarray}
The commutation relation of \(Q\) and \(P\) is obtained as   
\begin{eqnarray}
PQ = \omega QP \label{Wigner2}.
\end{eqnarray}
When the above Weyl operator is expressed using the phase and shift operators in \(I\) indexing,
it is given by
\begin{eqnarray}
{W^C}_{m,n}=\omega^{-2mn}Q^{2n}P^{-2m}\label{Wigner3}
.
\end{eqnarray}

\subsection{Definition of phase point operator in odd-lattice case}

Cohendet et al. define the phase point operator as the \(T\) transformation of the Weyl operator,
\begin{eqnarray}
{\Delta^C}_{m,n} = {W^C}_{m,n}T = \omega^{-2mn}Q^{2n}P^{-2m}T\label{Wigner4}
,
\end{eqnarray}
with the following properties:
\begin{eqnarray}
{{\Delta}^C}^{\dag}_{m,n} = {{\Delta}^C}_{m,n} 
,
\end{eqnarray}
\begin{eqnarray}
{\rm Tr}({{\Delta}^{C}}_{m,n}) = 1
,
\end{eqnarray}
\begin{eqnarray}
{\rm Tr}({{\Delta}^{C}}^{\dag}_{m,n}{\Delta^C}_{m',n'}) = N\delta_{m,m'}\delta_{n,n'} 
,
\end{eqnarray}
\begin{eqnarray}
{W^C}^{\dag}_{m',n'}{\Delta^C}_{m,n}{W^C}_{m',n'} = {\Delta^C}_{m-2m',n-2n'} 
.
\end{eqnarray}
The Hermiticity Eq. (18), normalization Eq. (19), traciality Eq. (20) and covariance Eq. (21)
are properties of the Stratonovich-Weyl kernel, which show the eligibility of the definition.


%
Using the phase point operator defined in Eq. (17), the Wigner function on odd-lattice phase space
is defined in the same way as in the continuous case Eq. (4), i.e., as
\begin{eqnarray}
{\cal W}_{m,n}
=
{1\over N}{\rm Tr}\left[ \rho{\Delta^C}_{m,n} \right]
=
{1\over N}\langle\psi|{\Delta^C}_{m,n}|\psi\rangle
.
\end{eqnarray}
The marginality of this Wigner function (discrete form of Eqs. (2) and (3))
and its operator form (discrete form of Eqs. (8) and (9)) can be confirmed in the same way as on continuous space. 

\section{Wigner function by Leonhardt and the phase point/Weyl operators on even-lattice phase space}

In this section,we review the Wigner function on discrete phase space
composed of an even number of lattice points (even-lattice phase space),
as described by Leonhardt \cite{Leonhardt1}.

\subsection{Definition of the Weyl operator in even-lattice case}

The Wigner function composed by Leonhardt is established on both odd- and even-lattice phase space,
but with incorporation of a virtual degree of freedom (ghost variable) between integral points for even-lattice phase space.
The 'characteristic function' is defined as 
\begin{eqnarray}
{{\tilde{W}}^L}_{m,n}\equiv \sum_{k=0}^{N-1} \exp \left[-\frac{2\pi i}{N}2m(k+n )\right] \langle k|\rho |k+2n  \rangle
.
\label{eq:leonhardt1}
\end{eqnarray}
Leonhardt defines the discrete Wigner function as a double-inverse Fourier transformation:
\begin{eqnarray}
{\cal W}_{\mu ,\nu  }\equiv \frac{1}{D^2}\sum_{m,n }\exp \left[ \frac{2\pi i}{N}2(m\nu +n \nu )\right] {{\tilde W}^L}_{m,n }
.
\label{eq:leonhardt2}
\end{eqnarray}
Substitution of Eq. (23) into Eq. (24) then yields
\begin{eqnarray}
{\cal W}_{\mu,\nu }=\frac{1}{D}\sum_{m}\exp \left( \frac{2\pi i}{N}2m\nu \right) \langle \mu -m|\rho| \mu +m \rangle
\label{eq:leonhardt3}
.
\end{eqnarray}
Eq. (25) is the Wigner function that has the marginality
on both odd- and even-lattice phase space,
provided that for odd dimensions \((\mu,\nu)\) is an integer phase space composed with \(D=N\)
and summed in the range
\(I=\{ -\frac{N-1}{2} ,-\frac{N-3}{2},\cdots ,\frac{N-3}{2},\frac{N-1}{2} \}\)
and that for even dimensions the phase space is composed with \(D=2N\)
together with \((\mu,\nu)\) integers and half-integers, summation is performed in the range
\(I'=\{ 0,\frac{1}{2},1, \cdots ,\frac{2N-1}{2}\}\) \(({\rm mod}\ N)\).
State vectors are set to zero on half-integer points.
The Wigner function is real and normalized to unity, i.e.,
\(\sum_{\mu,\nu}{\cal W}_{\mu,\nu}=1\).
As the characteristic function can be transformed to 
\begin{eqnarray}
{\tilde{W}^L}_{m, n }&=& \sum_{k=0}^{N-1} \omega^{-2m(k+n)}\langle \psi | k+2n \rangle \langle k | \psi \rangle \nonumber \\
&=&\langle \psi |{\omega}^{2mn}Q^{-2m}P^{-2n}|\psi \rangle \label{eq:leonhardt4}
,
\end{eqnarray}
and thus with
\({\tilde{W}^L}_{m, n }={\rm Tr}[\rho {W^L}_{m,n}]\)
the Weyl operator is then defined as
\begin{eqnarray}
{W^L}_{m,n}={\omega}^{2mn}Q^{-2m}P^{-2n}
.
\end{eqnarray}

\subsection{Definition of phase point operator in even-lattice case}

The Leonhardt phase point operator is defined by the double-inverse Fourier transformation
of \(W^L_{m,n}\),
\begin{eqnarray}
{\Delta^L}_{m,n}=\frac{1}{N}\sum_{m',n'\in I'}
\exp \left[ \frac{2\pi i}{N}2(mm'+nn')\right]   {W^L}_{m',n'}\label{eq:leonhardt4.5}
.
\end{eqnarray}
With the phase, shift, and inversion operator defined by Eqs. (11), (12), and (13)
indexed by
\(I''=\{ 0,1,\cdots,N-1\}\),
Eq. (28) can then be expressed as 
\begin{eqnarray}
{\Delta ^L}_{m,n}=
\omega ^{-2mn}Q^{2n}P^{-2m}T \ \ \ ,\ \ \ (m,n\in I')\label{eq:leonhardt5}
.
\end{eqnarray}
Eq. (29) reduce to Eq. (17) for odd \(N\) \((m,n\in I)\).

\section{Covariance relation}

\subsection{Translational covariance on continuous space and odd-lattice phase space}

When the phase point operator defined on continuous space
or the discrete phase point operator defined by Cohendet et al.
is sandwiched between Weyl operator and its Hermitian conjugate,
it has the property of determining the phase point operator at a
given point.

Using the continuous Weyl operator
\begin{eqnarray}
W(q,p)=\exp \left\{ \frac{i}{\hbar}(\hat{q}p-\hat{p}q)\right\} \label{continuous1}
,
\end{eqnarray}
the phase point operator can then be expressed as
\begin{eqnarray}
\Delta(q,p)=W(q,p)\Delta(0,0)W^{\dag}(q,p)
.
\end{eqnarray}
This shows that the Weyl operator is the projective representation of the group
representing translation of continuous phase space.
On the odd-lattice phase space, the Weyl operator is given by 
\begin{eqnarray}
W_{m,n}=\omega^{-mn/2}Q^{n}P^{-m}\label{continuous2}
,
\end{eqnarray}
the phase point operator can then be expressed as
\begin{eqnarray}
\Delta_{m,n}=W_{m,n}\Delta_{0,0}{W^\dag}_{m,n}\label{Covariance4}
.
\end{eqnarray}
Eqs. (32) and (33) are natural discretization of Eqs. (30) and (31).

On even-lattice phase space,
the projective representation of groups representing translation
remains unknown because of its inclusion
of the virtual degree of freedom (ghost variable).

\subsection{Symplectic covariance}

The symplectic transformation is a linear canonical transformation
on continuous phase space
that plays an essential role in classical mechanics
in keeping the Poisson bracket invariant.
It also provides an important symmetry in quantum mechanics
as a transformation rendering canonical commutation relation invariant.
%
The transformations form a symplectic group denoted as \(Sp(M,\mathbb R)\),
where \(M\) is the dimension of phase space.
For two dimensional case, it is given by
\begin{eqnarray}
Sp(2,\mathbb R)
=
\left \{
S=
\left(
\begin{array}{cc}
  a & b \\
  c & d
\end{array}
\right)
\Bigg|
\ a,\ b,\ c,\ d\in \mathbb R,\ 
{\rm det}\:S= 1
\right \}.
\label{symplecticgroup}
\end{eqnarray}
The covariance of the phase point operator \(\Delta(q,p)\)
under \(S\) is defined as
\begin{eqnarray}
U(S)\Delta(q,p)U^{\dag}(S)=\Delta(S\cdot(q,p))\label{Symplectic2}
\end{eqnarray}
where \(U(S)\) is a projective unitary representation of the symplectic group
and \(S\cdot(q,p)\) is an abbreviation of \(S\left(\begin{array}{c}q\\p\\\end{array}\right)\).
If Eq. (35) is established for all symplectic transformations \(S\),
similar to the above translational covariance,
then the phase point operator \(\Delta(q,p)\) is said to possess symplectic covariance.
On the discrete phase space, the covariance is defined in a similar way
(see the next section).
In previous studies, it was shown that the phase point operator
is uniquely determined under certain symplectic covariance \cite{Takami,Horibe1,Horibe2}.
%
%
%
However, the specific form of \(U(S)\) has not been explicitly described.

\section{Group of symplectic transformations}

\subsection{Definition and its generator}

We define the group \(Sp_N\) of symplectic transformations \(S\)
on the discrete phase space \(\mathbb Z_N\times\mathbb Z_N\)
by analogy with the continuous case Eq. (34),
\begin{eqnarray}
Sp_N=
\left \{
S=
\left(
\begin{array}{cc}
  a & b \\
  c & d
\end{array}
\right)
\Bigg|
\ a,\ b,\ c,\ d\in \mathbb Z_N,\ 
{\rm det}\:S= 1\in\mathbb Z_N
\right \},
\label{symplecticgroup}
\end{eqnarray}
where \(\mathbb Z_N\) is a residue ring modulo \(N\)
and its representatives are chosen from \(\{0,\cdots,N-1\}\).
The covariance relation Eq. (35) becomes
\begin{eqnarray}
U(S)\Delta_{m,n}U^{\dag}(S)=\Delta_{S\cdot(m,n)}
,
\end{eqnarray}
in this discrete case.

Here we show that the group \(Sp_N\) 
is generated from the two elements \(h_+\) and \(h_-\),
which are defined as
\begin{eqnarray}
h_+
{\equiv}
\left(
\begin{array}{cc}
 1 & 1 \\
 0 & 1 \\
\end{array}
\right){\in}Sp_N
,
\hspace{7mm}
h_-
{\equiv}
\left(
\begin{array}{cc}
 1 & 0 \\
 1 & 1 \\
\end{array}
\right){\in}Sp_N
.
\end{eqnarray}
We denote the group generated by \(h_+, h_-\) as \(Sp'\)
and have \(h_+^{-1}=h_+^{N-1}, h_-^{-1}=h_-^{N-1}\),
as \(h_+^N=h_-^N=I\).

Let \(S\) be an arbitrary element in \(Sp_N\).
Multiplying \(S\) by \(h_+\) and \(h_-\), we obtain 
\begin{eqnarray}
h_+^nS
&=&
\left(
\begin{array}{cc}
 a+nc & b+nd \\
 c & d \\
\end{array}
\right),\label{EA1}
\\
Sh_+^n
&=&
\left(
\begin{array}{cc}
 a & na+b \\
 c & nc+d \\
\end{array}
\right),\label{EA2}
\\
h_-^nS
&=&
\left(
\begin{array}{cc}
 a & b \\
 na+c & nb+d \\
\end{array}
\right),\label{EA3}
\\
Sh_-^n
&=&
\left(
\begin{array}{cc}
 a+nb & b \\
 c+nd & d \\
\end{array}
\right),\label{EA4}
\end{eqnarray}
in which we perform the operation of
multiplying a row (column) by an element in \(\mathbb Z_N\)
and then adding the result to the other row (column).
We next define \(h_t\) as 
\begin{eqnarray}
h_t
{\equiv}
\left(
\begin{array}{cc}
 0 & 1 \\
 N-1 & 0 \\
\end{array}
\right){\in}Sp_N,
\end{eqnarray}
which can be represented in a form having \(h_+\) on both sides of \(h_-^{N-1}\),
\begin{eqnarray}
h_t
=
h_+h_-^{N-1}h_+
.\label{even5}
\end{eqnarray}
Multiplying \(S\) by \(h_t\), we then have
\begin{eqnarray}
h_tS&=&
\left(
\begin{array}{cc}
 c & d \\
 N-a & N-b \\
\end{array}
\right)
,\label{EA5}\\
Sh_t&=&
\left(
\begin{array}{cc}
 N-b & a \\
 N-d & c \\
\end{array}
\right)
.\label{EA6}
\end{eqnarray}
We have thus performed the operation of interchanging rows and columns.
Hence multiplying \(S\) by \(h_+\) and \(h_-\) on the left and right appropriately,
a given symplectic transformation \(S\in Sp_N\) can be transformed into \(h_t\).
This means that \(S\) can be
represented by \(h_+\) and \(h_-\), i.e.,
\begin{eqnarray}
S=\prod_ih_{s_i}
,
\hspace{10mm}
s_i\in\{+,-\}
\end{eqnarray}
and
\begin{eqnarray}
Sp_N=Sp'_N
.
\end{eqnarray}

The explicit procedure is given as follows.
We denote the Euclidean algorithm for \(b\) and \(d\) as
\begin{eqnarray}
r_0=\max(b,d),
\hspace{10mm}
r_1=\min(b,d),
\end{eqnarray}
\begin{eqnarray}
r_i=k_ir_{i+1}+r_{i+2}
\hspace{10mm}
(r_{i+2}< r_{i+1}; i=0,\cdots,l-2; l\ge2),
\end{eqnarray}
\begin{eqnarray}
r_l=0.
\end{eqnarray}
If \(b=d\) the procedure stops at the first step with \(l=2\).
Multiplying \(S\) by \(H\) defined as
\begin{eqnarray}
H=h_+^{-k_{l-1}}\cdots h_-^{-k_1}h_+^{-k_0}
,
\hspace{10mm}
{\rm for}
\ \ 
b>d
,
\ l:{\rm odd}
,
\end{eqnarray}
\begin{eqnarray}
H=h_th_-^{-k_{l-1}}\cdots h_-^{-k_1}h_+^{-k_0}
,
\hspace{10mm}
{\rm for}
\ \ 
b>d
,
\ l:{\rm even}
,
\end{eqnarray}
\begin{eqnarray}
H=h_+^{-k_{l-1}}\cdots h_-^{-k_1}h_+^{-k_0}h_t
,
\hspace{10mm}
{\rm for}
\ \ 
b<d
,
\ l:{\rm odd}
,
\end{eqnarray}
\begin{eqnarray}
H=h_th_-^{-k_{l-1}}\cdots h_-^{-k_1}h_+^{-k_0}h_t
,
\hspace{10mm}
{\rm for}
\ \ 
b<d
,
\ l:{\rm even}
,
\end{eqnarray}
from left,
\(S\) can be transformed into
\begin{eqnarray}
HS
=
\left(
\begin{array}{cc}
\alpha & 0 \\
\gamma & \beta \\
\end{array}
\right)
,
\hspace{10mm}
\alpha\beta=1\mod N.
\end{eqnarray}
Multiplying the left-hand side of Eq. (56)
by \(h_+^\beta h_-^{-\alpha}\) from the right
and by \(h_-^{-\beta-\beta\gamma}\) from the left, we have
\begin{eqnarray}
h_-^{-\beta-\beta\gamma}
\ HS\ 
h_+^\beta h_-^{-\alpha}
=
h_t
.
\end{eqnarray}
This proves Eq. (48).


\section{Uniqueness and explicit form of the projective representation}

Let us now show that the unitary representation \(U(S)\) of \(Sp_N\)
having the covariance Eq. (37) is determined up to a phase factor,
hence its projective representation is unique.
We also derive its explicit form using the Euclidean algorithm in this section.

\subsection{Uniqueness and explicit form of the representation}

Multiplying a new \(U(S')\) and its Hermitian conjugate from both side in Eq. (37), we can transform it to
\begin{eqnarray}
U(S')(U(S) \Delta_{m,n} U^{\dagger } (S))U^{\dagger }(S') 
= \Delta_{S'\cdot(S\cdot(m,n))}
.
\end{eqnarray}
Taking \(S'S=S''\), we have by definition
\begin{eqnarray}
U(S") \Delta_{m,n} U^{\dagger } (S") = \Delta_{S"\cdot(m,n)}
,
\end{eqnarray}
hence,
\begin{eqnarray}
(U(S')U(S))\Delta_{m,n}(U(S')U(S))^\dagger
=
U(S") \Delta_{m,n} U^{\dagger } (S")
.
\end{eqnarray}
From the traciality Eq. (20), the operators that commute with all phase point operators
are phase factor multiples of unit operator, and we therefore have
\begin{eqnarray}
U(SS')=\mathrm{e}^{i\theta}U(S)U(S'),\label{Shadow1}
\end{eqnarray}
thus showing that \(U(S)\) satisfying Eq. (37) is
a unitary projective representation of \(Sp_N\).

Let \(U'(S)\) be another such representation.
In Eq. (37), if we multiply the both sides by \(U'^\dagger(S)\) from the left
and by \(U'(S)\) from the right, we get
\begin{eqnarray}
(U'^\dagger(S)U(S))\Delta_{m,n}(U^\dagger(S)U'(S))=\Delta_{S^{-1}\cdot(S\cdot(m,n))}=\Delta_{m,n}.
\end{eqnarray}
Hence, using the traciality again, we have
\begin{eqnarray}
U'(S)=e^{i\theta}U(S).
\end{eqnarray}
Thus, the projective representation is unique.
%


Considering in conjunction with the uniqueness that a given symplectic transformation \(S\)
is represented by \(h_+\) and \(h_-\),
we find that a given \(U(S)\) can be represented by \(U(h_+)\) and \(U(h_-)\) as
\begin{eqnarray}
U(S)=\prod_iU(h_{s_i}),
\hspace{10mm}
s_i\in\{+,-\}.
\end{eqnarray}
The sign factor \(s_i\) is determined from the Euclidean algorithm.
One of such examples is given by
\begin{eqnarray}
U(S)=U^{-1}(H)U^{-\beta-\beta\gamma}(h_-)U(h_t)U^\alpha(h_-)U^{-\beta}(h_+)
,
\end{eqnarray}
using Eqs. (56), (57), where \(U(H)\) is the product of \(U(h_+)\) and \(U(h_-)\)
according to Eqs. (52)-(55) in the previous section, and \(U(h_t)=U(h_+)U^{-1}(h_-)U(h_+)\) using Eq. (44).

In Secs.VIII and IX, we consider \(U(h_+)\) and \(U(h_-)\) in more detail.

\section{\(U(h_+)\) and \(U(h_-)\) on odd-lattice phase space}

\subsection{Derivation of the explicit form \(U(h+)\) and \(U(h-)\)}
Multiplying both sides of Eq. (37) by \(U(S)\) from the right with \(S=h_\pm\) , we have
\begin{eqnarray}
U(h_\pm ){\Delta^C}_{m,n}={\Delta^C}_{h_\pm\cdot(m,n)}U(h_\pm ).\label{Symplectic3}
\end{eqnarray}
We use the phase point operators described earlier in Eq. (17)
to find an explicit form of \(U(h_+)\) and \(U(h_-)\).
We performed the actual calculation for lower dimensions (e.g. \(N=3,5,7\)).
For example, we have for \(N=7\)
\begin{eqnarray}
U(h_+)=\frac{1}{\sqrt{7}}
\left(
\begin{array}{ccccccc}
1 & \omega^4 & \omega^2 & \omega & \omega & \omega^2 & \omega^4 \\
\omega^4 & 1 & \omega^4 & \omega^2 & \omega & \omega & \omega^2 \\
\omega^2 & \omega^4 & 1 & \omega^4 & \omega^2 & \omega & \omega \\
\omega & \omega^2 & \omega^4 & 1 & \omega^4 & \omega^2 & \omega \\
\omega & \omega & \omega^2 & \omega^4 & 1 & \omega^4 & \omega^2 \\
\omega^2 & \omega & \omega & \omega^2 & \omega^4 & 1 & \omega^4 \\
\omega^4 & \omega^2 & \omega & \omega & \omega^2 & \omega^4 & 1 \\
\end{array}
\right)
,
\end{eqnarray}
\begin{eqnarray}
U(h_-)=
\left(
\begin{array}{ccccccc}
\omega & & & & & & \\
& \omega^2 & & & & & \\
& & \omega^4 & & & & \\
& & & 1 & & & \\
& & & & \omega^4 & & \\
& & & & & \omega^2 & \\
& & & & & & \omega \\
\end{array}
\right)
.\end{eqnarray}
The obtained results suggest that the general form can be thus given for odd dimensions as
\begin{eqnarray}
U(h_+)&=&\frac{1}{\sqrt{N}} \sum_{i,k\in I}|i \rangle \omega^{\frac{1}{2}(i-k)(i-k+N)} \langle k| \label{U+}
,\\
U(h_-)&=&\sum_{i\in I} |i \rangle {\omega}^{\frac{1}{2} i(i+N)} \langle i| \label{expect1}
.
\end{eqnarray}
%

%
Let us now confirm the symplectic covariance
of the predicted \(U(h_+)\) and \(U(h_-)\) for a given phase point operator
\(\Delta_{m,n}^C\).
In bra-ket notation, the phase point operator in Eq. (17) \(\Delta_{m,n}^C\) is
\begin{eqnarray}
{\Delta^C}_{m,n}=
\sum_{i}\omega^{2n(-m+i)}| i \rangle   \langle -i+2m | 
.
\end{eqnarray}
Accordingly,
\begin{eqnarray}
U(h_+){\Delta^C}_{m,n}U^\dag(h_+)\nonumber
\end{eqnarray}
\vspace{-20pt}
\begin{eqnarray}
&=&
\sum_{i}\omega^{2n(-(m+n)+i)}|i\rangle\langle-i+2(m+n)| \nonumber \\
&=&
{\Delta^C}_{m+n,n}
=
{\Delta^C}_{h_+\cdot(m,n)}
,
\end{eqnarray}
thus confirming the symplectic covariance.
The general form of \(U(h_+)\) for odd dimensions
can therefore be regarded as Eq. (69).
%
For \(U(h_-)\), the symplectic covariance of Eq. (70)
is similarly confirmed.
%

\section{\(U(h_+)\) and \(U(h_-)\) on even-lattice phase space}

\subsection{Extension of the dimension of symplectic group}

We derive \(U(h_+)\) and \(U(h_-)\) on even-lattice phase space
in the same manner as the above derivation on odd-lattice phase space.
More specifically, using Leonhardt's phase point operator with \(N=2\), \(U(h_+)\) and \(U(h_-)\)
having symplectic covariance are
\begin{eqnarray}
U(h_+)&=&\frac{1}{\sqrt{2}}
\left(
\begin{array}{cc}
1 & i\\
i & 1 \\
\end{array}
\right),\label{even1} \\
U(h_-)&=&
\left(
\begin{array}{cc}
1 & 0 \\
0 & i  \\
\end{array}
\right).\label{even1.5}
\end{eqnarray}
However, these two explicit forms are not a projective representation
of the symplectic group as defined by Eq. (36).
To compose the Wigner function on discrete phase space in even dimensions,
Leonhardt incorporated a virtual degree of freedom (ghost variable)
and multiplied the number of variables by two.
It is therefore also necessary to reconsider operation of the symplectic groups
on even-lattice phase space.
In the following, we redefine the symplectic group in even dimensions. 
%
%
%
%
An even-lattice phase space is a \(\mathbb Z_{2N}\times\mathbb Z_{2N}\) space
taking into the ghost degree of freedom into account,
and the symplectic group in this case is defined as        
\begin{eqnarray}
Sp_{2N}=
\left \{
\left(
\begin{array}{cc}
  a & b \\
  c & d
\end{array}
\right)
\Bigg|
\ a,\ b,\ c,\ d\in \mathbb Z_{2N},\ 
{\rm det}\:S=1\in\mathbb Z_{2N}
\right \} 
\end{eqnarray}
We consider the projective representation based on this definition.

\subsection{Derivation of the explicit form \(U(h+)\) and \(U(h-)\)}

From Eq. (73), we now have
\begin{eqnarray}
\left\{ U(h_+) \right\}^4 
&=&
\frac{1}{4}
\left(
\begin{array}{cc}
-4 & 0 \\
0 & -4 \\
\end{array}
\right)
\doteq
\left(
\begin{array}{cc}
1 & 0 \\
0 & 1 \\
\end{array}
\right)
\hspace{5mm}
{\rm (up\ to\ a\ phase\ factor)} \nonumber \\
&=& U(E)=U({h_+}^4 )
\end{eqnarray}
where \(E\) is the unit element of \(Sp_{2N}\),
thereby confirming at least one projective representation of \(h_+\in Sp_{2N}\).  
From \(U(h_+)\) and \(U(h_-)\) at \(N=2\) and from other lower dimensional examples,
we predict 
\begin{eqnarray}
U(h_+)&=&\frac{1}{\sqrt{N}} \sum_{i,k\in I''}|i \rangle \tilde{\omega}^{(i-k)^2} \langle k|,\label{even3}\\
U(h_-)&=&\sum_{i\in I''} |i \rangle {\tilde{\omega}}^{i^2} \langle i|,\label{even2}
\end{eqnarray}
where \(\tilde\omega=\omega^{\frac12}=+\exp\left(\frac{2\pi i}{2N}\right)\)
and \(I''=\{ 0,1,\cdots,N-1\}\).
%


%
In the same manner as for odd dimensions, we next confirm
the symplectic covariance by the predicted \(U(h_+)\) and \(U(h_-)\):
\begin{eqnarray}
U(h_\pm ){\Delta^L}_{m,n}U^\dag (h_\pm )={\Delta^L}_{h_\pm\cdot(m,n)},
\end{eqnarray}
for the phase point operator \(\Delta_{m,n}^L\) defined by Leonhardt (Eq. (29)):
\begin{eqnarray}
{\Delta^L}_{m,n}&=&\tilde{\omega}^{-4mn}Q^{2n}P^{-2m}T \nonumber \\
&=&\sum_{i\in I''}\tilde{\omega}^{4n(i-m)} |i\rangle \langle -i+2m |.
\end{eqnarray} 
Note that \(m,n\in I'\).
%
%
For \(U(h-)\), we obtain 
\begin{eqnarray}
U(h_-){\Delta^L}_{m,n}U^\dag (h_-)  \nonumber 
\end{eqnarray}
\begin{eqnarray}
&=&
\sum_{i\in I''}\tilde{\omega}^{4(m+n)(i-m)}| i \rangle   \langle -i+2m | \nonumber \\
&=&
{\Delta^L}_{m,m+n}
=
{\Delta^L}_{h_-\cdot(m,n)},
\end{eqnarray}
thereby confirming symplectic covariance for \(Sp_{2N}\)
as the group of symplectic transformations \(S\).
Accordingly, the form of \(U(h_-)\) in even dimensions in general can be regarded as 
in Eq. (78).
%
%
For \(U(h_+)\), we can similarly confirm that Eq. (77) has symplectic covariance
with the Leonhardt's phase point operator.

\section{Summary and conclusion}

In the present study, we have determined the projective unitary representation
\(U(S)\) of the discrete symplectic group \(Sp_N\) and \(Sp_{2N}\)
on odd-lattice phase space and on even-lattice phase space, respectively,
which satisfy symplectic covariance for the phase point operator \(\Delta_{m,n}\) (Eq. (37)):
\[
U(S)\Delta_{m,n}U^{\dag}(S)=\Delta_{S\cdot(m,n)}
.
\]
The phase point operator defined by Cohendet et al. is used for the odd case,
and that by Leonhardt for the even case
to search for the representations.

A symplectic transformation \(S\in Sp_N\) can be obtained
by the product of a series of \(h_+\) and \(h_-\) (Eq. (38)):
\[
h_+
{\equiv}
\left(
\begin{array}{cc}
 1 & 1 \\
 0 & 1 \\
\end{array}
\right){\in}Sp_N
,
\hspace{7mm}
h_-
{\equiv}
\left(
\begin{array}{cc}
 1 & 0 \\
 1 & 1 \\
\end{array}
\right){\in}Sp_N
,
\]
as in Eq. (47):
\[
S=\prod_ih_{s_i},
\hspace{10mm}
s_i\in\{ +,-\}
,
\]
using the Euclidean algorithm.
As the projective unitary representation satisfying the covariance Eq. (37) is unique,
an explicit form of \(U(S)\) is given by the product of the same series,
replacing \(h_+\) and \(h_-\) by \(U(h_+)\) and \(U(h_-)\), respectively,
\[
U(S)=\prod_iU(h_{s_i}),
\hspace{10mm}
s_i\in\{+,-\}
\]
(See Eq. (64)).
One of these explicit forms is shown in Eq. (65).
This holds also for \(Sp_{2N}\).

The unitary operators \(U(h_+)\) and \(U(h_-)\) for \(Sp_N\) are given by Eqs. (69) and (70):
\[
U(h_+)_{\rm{odd}}=\sum_{i,k\in I}|i \rangle \omega^{\frac{1}{2}(i-k)(i-k+N)} \langle k|
,
\]
\[
U(h_-)_{\rm{odd}}=\sum_{i\in I} | i \rangle \omega^{\frac{1}{2}i(i+N)} \langle i|
,
\]
on odd-lattice phase space, and that for \(Sp_{2N}\) by Eqs.(77) and (78):
\[
U(h_+)_{\rm{even}}=\sum_{i,k\in I''}|i \rangle \tilde{\omega}^{(i-k)^2} \langle k|
,
\]
\[
U(h_-)_{\rm{even}}=\sum_{i\in I''} | i \rangle \tilde{\omega}^{i^2} \langle i|
,
\]
on even-lattice phase space.
The results thus show that a projective unitary representation
of the symplectic transformation groups that satisfies the covariance relation
exists and is unique,
and that its specific \(U(S)\) forms can be derived using the Euclidean algorithm.
A question inviting further study is whether a projective representation of groups
similarly having symplectic covariance uniquely exists on continuous space,
as it does on odd- and even-lattice phase space.
%

%
For discrete spaces, straightforward canonical quantization cannot be performed
by replacing the momentum operator \(\hat p\) by the differential operator \(-ih\frac\partial{\partial q}\).
%
If we multiply the Hamiltonian by the phase point operator and integrate, however,
it is possible to obtain its canonical quantization with Weyl ordering.
%
If the phase point operator can be found, then quantization becomes possible.
This has been proved for cases on the continuous phase space or on odd-lattice phase space.
For application to discrete phase space composed of even-dimensional lattice points,
unique determination of the phase point operator is necessary.

\end{document}